\documentstyle[preprint,eqsecnum,aps]{revtex}
\draft 
\begin{document}

%\twocolumn [
%\hsize\textwidth\columnwidth\hsize\csname @twocolumnfalse\endcsname

\date{\today} 
\title{Structure-Factor Tail for the Ordering Kinetics of Non-conserved 
Systems without Topological Defects}
\author{F. Rojas and A. J. Bray}
  
\address{Theoretical Physics Group \\
         Department of Physics and Astronomy \\
         The University of Manchester, M13 9PL, UK}

\maketitle  
%\widetext

\begin{abstract}
Using a cell dynamic system (CDS) simulation scheme, we investigate the 
phase-ordering dynamics of non-conserved $O(n)$ models without topological 
defects, i.e.\ for $n > d+1$ where $d$ is the spatial dimensionality.  
In particular, we  consider zero-temperature quenches for $d=2$, $n=4,5$, 
and for $d=1$, $n=3,4,5$. We find, in agreement with previous simulations 
using fixed-length spins,  that dynamical scaling is obtained, with 
characteristic length $L(t) = t^{1/2}$. We show that the asymptotic 
behaviour of  the structure-factor scaling function $g(q)$ is well fitted 
by the stretched exponential form $g(q)\sim \exp(-bq^\delta)$, with an 
exponent $\delta$ that appears to depend on both $n$ and $d$. 
An analytical treatment of an approximate large-$n$ equation for the pair 
correlation function yields $g(q) \sim q^{-(d-1)/2}\exp(-bq)$, 
with $b \sim (\ln n)^{1/2}$ for large $n$, in agreement with recent 
simulations of the same equation.  
\end{abstract}

\medskip
\narrowtext 
%]

\section{INTRODUCTION}
The phase-ordering dynamics of systems quenched from a high-temperature 
disordered state into an ordered state is a problem of great relevance in 
the description of out-of-equilibrium pattern formation \cite{REV}.  
One well established property is the onset of dynamic scaling, 
where the late-time behaviour of the order-parameter correlation 
functions is described by scaling forms with a single time-dependent 
length scale $L(t)$. Thus the real-space correlation function is found to 
have the scaling form 
\begin{equation}
C(r,t)=  f(r/L(t))\ ,
\label{scaling:real}
\end{equation}
while its Fourier transform, the structure factor, has the 
corresponding scaling form
\begin{equation}
S(k,t)= [L(t)]^d g(kL(t))\ .
\label{scaling:Fourier}
\end{equation}

Conventional experimental systems such as binary alloys and binary liquids 
are described by a scalar order parameter. Recently, however, there has 
been much interest in systems with more complicated order parameters such 
as $n$-component vectors (the $O(n)$ model) \cite{REV}--\cite{TU}
and traceless symmetric tensors (nematic liquid crystals) \cite{Nematics}. 
In this paper we restrict discussion to the $O(n)$ model. 

Much numerical and theoretical effort has been devoted to 
understanding the basic properties of systems that can support singular 
topological defects, i.e.\ systems with $n \le d$. The presence 
of such defects leads to a generalization of the usual Porod law for the 
large-$q$ tail of the structure-factor scaling function $g(q)$, namely
$g(q) \sim q^{-(d+n)}$ for $q \gg 1$ \cite{Porod}. 
This result is geometrical in origin, and is independent of whether or 
not the order parameter is conserved by the dynamics \cite{BH}. 

Very recently, the cases $n=d+1$, for which nonsingular topological 
textures occur, have been studied numerically (for $d=2$) and 
analytically (for $d=1$). Rutenberg and Bray \cite{OXY}  found  that the 
$d=1$  XY model ($n=2$) exhibits a scaling violation due to the existence 
of {\em two} relevant length scales: the phase coherence length and phase 
winding length. On the other hand, the two-dimensional Heisenberg model 
($n=3$)  with non-conserved dynamics also violates  dynamic scaling due, 
it appears, to the existence of as many as three separate length scales, 
related to individual texture size, the typical separation between textures 
and the typical distance between textures of opposite charge \cite{TOT}. 
These texture systems are, perhaps, the most complex of the phase 
ordering systems.
 
By contrast, systems without topological defects ($n>d+1$) seem relatively 
straightforward. There is good evidence for the simple scaling behavior 
described by (\ref{scaling:real}) and (\ref{scaling:Fourier}), with
characteristic scale $L(t) \sim t^{1/2}$ for nonconserved dynamics. 
The energy scaling approach of Bray and Rutenberg \cite{AR} shows that, 
provided scaling holds, $L(t) \sim t^{1/2}$ is indeed correct for $n>d+1$ 
systems with nonconserved dynamics, and gives $L(t) \sim t^{1/4}$ for 
conserved dynamics, again nicely consistent with simulation results \cite{RC} 
and the Renormalization Group result of Bray \cite{BREN}. 
 
Much less is known, however, about the form of the 
structure-factor tail for $n>d+1$. The recent simulations of Rao and 
Chakrabarty (RC) \cite{RC}, with conserved dynamics, for the cases  
$d=1$, $n=3$ and $d=2$, $n=4$ show `squeezed exponential' behavior 
[i.e.\ $g(q) \sim \exp(-bq^\delta)$ with $\delta>1$].  In this paper 
we concentrate on systems with $n>d+1$ and nonconserved dynamics. We 
consider the cases $d=2$, $n=4,5$ and $d=1$, $n=3,4,5$. 
In each case we confirm the expected $t^{1/2}$ growth, and 
find `stretched exponential' behavior [i.e.\ $g(q) \sim \exp(-bq^\delta)$ 
with $\delta \le 1$] for the tail of the structure factor, with an exponent 
$\delta$ that appears to depend on $n$ and $d$. 

In an attempt to understand the origin of this tail behaviour, we present 
an analytical approach based on an approximate equation due to Bray and 
Humayun (BH) \cite{BH92}, which is itself based on the `gaussian auxiliary 
field' (GAF) method \cite{GAF} that describes rather well the form of the 
structure factor for nonconserved systems with singular defects ($n\le d$).  
For those systems, this method reproduces, in particular, the generalized 
Porod tail. For $n>d+1$, the physical basis of the method is less clear. 
However, the simple truncation of the equation at leading order in $1/n$, 
proposed by BH in another context \cite{BH92}, leads to an exponential decay 
of $g(q)$, modified by a power-law prefactor for $d>1$. It is noteworthy 
that the asymptotic behaviour is nonanalytic in $1/n$: for $n$ strictly
infinite, the gaussian form $g(q) \sim \exp(-2q^2)$ is obtained. The 
exponential asymptotics of the BH equation were noted in recent numerical 
studies by Castellano and Zannetti \cite{Cast}. 

Using a `hard-spin' model Newman et al.\ \cite{NEW} studied numerically 
the dynamics  of one-dimensional systems without defects for $n=3,4$ and $5$. 
Measuring only the real-space correlations, they found that dynamic  scaling 
is obeyed with characteristic length  $L(t)=t^{1/2}$. Moreover, they found 
the real-space correlation function was very well fitted by a gaussian form,  
which is the exact result in the limit $n \to \infty$. 
The Fourier space analysis presented here, revealing stretched exponential 
tails, shows that the good gaussian fits achieved in real space are 
misleading.        

Our main results can be summarized as follows: (i) For all our models the 
characteristic length scale required to collapse the data for the 
real-space correlation function and structure factor, is $L(t)= t^{1/2}$,    
in agreement with theoretical predictions \cite{AR}; 
(ii) The asymptotic behaviour of the structure factor is well described 
by a stretched exponential of the form 
$g(q) \sim \exp(-bq^\delta)$, where the exponent $\delta$ apparently 
depends on both $n$ and $d$ and seems to be different from the value obtained 
for the corresponding system with conserved dynamics \cite{RC};  
(iii) An analytical treatment of the approximate BH equation, expected to 
be valid at large (but finite) $n$, gives a simple exponential modified by 
a power, $g(q) \sim q^{-(d-1)/2}\exp(-bq)$, with the {\em same} asymptotic 
form for conserved dynamics.  

The rest of the article is organized as follows. 
In the next section, we introduce the CDS model based on the time-dependent 
Ginzburg-Landau (TDGL) equation for a zero temperature quench, and  we 
describe the corresponding numerical procedure employed in the simulation.
Section 3 presents simulation results for a vector order
parameter with $n=4$ and $n=5$ components in one and two dimensions
and the one-dimensional $O(3)$ model. For the $d=2$ systems, we also 
present data for the real-space correlation function to demonstrate the 
dynamic scaling behaviour. We then discuss the procedure used to obtain 
the asymptotic functional form of the structure factor tail. 
Next we compare the data to the results of the approximate analytic 
theory. Finally we make some concluding comments on our results and a 
give a brief summary.

\section{MODEL  AND SIMULATIONS}
The dynamic evolution of a non-conserved vector order parameter
(model A) with $n$ components $\vec{\phi} = (\phi_1,\phi_2,\cdots,\phi_n)$,
for a zero-temperature quench, is described by a purely dissipative process 
defined in term of the following TDGL equation: 
\begin{equation}
\frac{\partial \vec{\phi}({\bf x},t)}{\partial t} = - \Gamma \frac{\delta 
F(\vec{\phi}({\bf x},t))}{\delta \vec{\phi}({\bf x},t)}\ ,  \label{eq:smai} 
\label{TDGL}
\end{equation}
with $\Gamma$ a kinetic coefficient that we will set equal to unity,
and $F$ the free energy functional which generates the thermodynamic force, 
\begin{equation}
F[\vec{\phi}({\bf x},t)] = \int  d^d{\bf x} \left[ \frac{1}{2} 
(\nabla \vec{\phi}({\bf x},t))^2 + V(\vec{\phi}({\bf x},t)) \right]\ ,   
\label{F}
\end{equation}
with the potential defined as
\begin{equation}
    V(\vec{\phi}({\bf x},t)) = \frac{1}{4}(1-\vec{\phi}^2({\bf x},t))^2
\end{equation}
where $\vec{\phi}^2 = \sum_{i=1}^{n} \phi^2_i ({\bf x},t)$.     
The ground states, or fixed points of the dynamics, are determined by the 
condition $\vec{\phi}^2 = 1$,  which defines a  degenerate manifold of states 
connected by rotations. In the internal space of the order parameter, 
this manifold is the surface of an $n$-dimensional sphere. At late times 
the order parameter is saturated in length (i.e.\ lies on the ground 
state manifold everywhere).  Then the dynamics is driven by the 
decrease of the free-energy associated with the term $(\nabla
\vec{\phi})^2$ in (\ref{F}), through a reduction in the magnitude of the 
spatial gradients.    
 
We can construct an explicit numerical scheme for the simulation based on 
a computationally efficient algorithm, namely the cell-dynamic system (CDS) 
\cite{OP1}, which updates the order parameter according to the rule
\begin{equation}
\vec{\phi}({\bf x},t+1) = H(\vec{\phi}({\bf x},t))  + \tau D \left[ \frac{1}{z}
 \sum_{\bf x'}\vec{\phi}({\bf x'},t) - \vec{\phi}({\bf x},t) \right]\ ,   
\end{equation}
with
\begin{equation}
H(\vec{\phi}({\bf x},t))= \vec{\phi}({\bf x},t) +\tau 
\vec{\phi}({\bf x},t)(1-(\vec{\phi}({\bf x},t)^2)\ ,  
\end {equation}
where $z$ is the number of nearest neighbors, and $\tau$ and $D$ are
parameters that we choose  to be  $\tau =0.2$ and $D=0.5$
in our simulations.

The above numerical procedure is identical to that used by
Toyoki \cite{TU}, differing only in the values of the  
parameters $\tau$ and $D$. The CDS is an Euler-like algorithm and 
for convenience in our analysis of the results we use a unit of time 
equal to the update time step $\tau$. It should be noted (see Figures 
1 and 3) that the scaling regime is reached very quickly in these 
systems without defects, and very long runs are not necessary. 

The two-dimensional systems consist of a square lattice of
size  $256\, \times \,256$  with periodic boundary conditions. 
The physical quantities are calculated as averages over 20 independent 
distributions of initial conditions. The one-dimensional systems  
have $L= 16384$ sites (with periodic boundary conditions) and we average 
$100$ independent runs. The initial conditions for the  order parameter 
components $\vec{\phi}_i$  were randomly chosen from a uniform distribution 
with support on the interval (-0.1,0.1).

A quantity of interest  that is computed during the course
of the numerical simulation in the two-dimensional models is
the two-point real-space correlation function
\begin{equation}
C({\bf r},t)= <\vec{\phi}({\bf x},t)\cdot\vec{\phi}({\bf x}+{\bf r},t)> 
\label{eq:cor} 
\end{equation}
where $<\cdots>$ stands for the average over the set of independent 
initial conditions (or `runs'). A spherical average over all possible 
distances $r=|{\bf r}|$ is performed to find the isotropic real-space 
correlation $C(r,t)$.  The other function of interest, calculated for 
all the models, is the structure factor, 
\begin{equation}
S({\bf k},t)= <\vec{\phi}({\bf k},t)\cdot\vec{\phi}({\bf -k},t)>\ .
\label{eq:struc}
\end{equation}
We also make a spherical average over all possible values of  
${\bf k}$ with given $k= |{\bf k}|$.
 
In the calculation of these quantities at each time, the data are 
`hardened' by replacing the order parameter at each point by a unit vector 
in the same direction (the fixed point of the CDS iteration being a vector 
of unit length). This procedure accelerates the entry into the dynamic 
scaling regime,  and helps us to elucidate the proper nature of the 
asymptotic tail in the structure factor.

\section{RESULTS} 
Dynamic scaling is observed for  all the models studied.   
The scaling regime is reached at quite early times, in agreement 
with previous studies. We show that dynamic scaling holds in the 
two-dimensional systems ($n=4,5$), using the characteristic length
$L(t) = t^{1/2}$ deduced from theoretical considerations \cite{AR}. 
This agrees with earlier simulations of Bray and Humayun using `hard-spin'
dynamics \cite{BH90}. 
 
Figure 1(a) presents a plot for $d=2$, $n=4$ of the correlation function 
(\ref{scaling:real}) as a function of distance $r$ for several 
times, while in Figure 1(b) we  show  the collapsed dynamic scaling 
function when the analysis is made using the
scaling variable $x = r/L(t)$. As can be seen from the Figure, 
the scaling function $f(x)$ is a monotonically decreasing function 
with the generic featureless shape that is characteristic of 
non-conserved $O(n)$ models. 

It is of some interest to investigate the small-$x$ behavior of the 
real-space scaling function $f(x)$. In systems with $n \le d$, the existence 
of singular topological defects leads to a non-analytic term of the form 
$|x|^n$ (with an additional $\ln x$ factor for even $n$), which leads to 
the $k^{-(d+n)}$ Porod tail in Fourier space \cite{Porod,BH}. 
In the present case, where 
$n>d$, we expect no such short-distance singularities. 
Therefore, we consider an expansion of the form 
$f(x) =  1- \alpha x^2 + \beta x^4 \cdots $.
In table 1 we present the parameters $\alpha$ and $\beta$ determined from 
the simulations in the range $x < 0.5$. The ratio 
$r=\beta/\alpha^2$ should be a universal number for given $n$ and $d$. 
It will be seen from table 1 that this ratio has the value 
$r \simeq 0.59$ for $n=4$, different from the value 
$1/2$ obtained for a gaussian function, which is the exact result 
for the limit $n \to \infty$. For $n=5$, table 1 gives $r \simeq 0.49$, 
already consistent with the large-$n$ result. However, in the absence of 
any short-distance singularity, the  small-$x$ behavior provides no useful 
information on the nature of the tail in  the structure factor. 
Consequently, it is more convenient to investigate 
directly the simulation results of the structure factor and
extract from them the asymptotic behaviour. We shall see that the 
behavior in Fourier space is clearly non-gaussian, even for $n=5$. 
 
As expected, given the absence of topological defects, the results 
indicate (Fig.\ 2) that the decay of the structure factor is clearly 
faster than a power law, in contrast to the interpretation of his own 
similar results by Toyoki \cite{TU}. In order to demonstrate that the 
tail is well described by the stretched exponential form 
\begin{equation}
g(q) \sim  A\exp(-bq^{\delta})\ ,
\label{se}
\end{equation} 
where $q=kL(t)$ is the scaling variable in momentum space, we attempt 
to find the corresponding power $\delta$ in the exponential by plotting 
$\ln g$ versus $q^\delta$ and adjusting the value of $\delta$ until 
the best linear behavior is obtained in the regime $q>1$. 
During the fitting procedure the other two parameters of the fit, 
$A$ and $b$, are readily determined.  
The criteria used for the optimum fitting is based on the 
Pearson correlation coefficient (PCC), which measure the strength of 
the linear relation among two variables and varies between -1 (perfect
negative linear relation) and $+1$ (perfect positive linear relationship).
We proceed as follows: first, we choose an  exponent $\delta$ and then 
perform  linear regression; next we change $\delta$ until the PCC 
reaches its maximum value. The regression coefficients are calculated 
using the values of the scaling structure function at the  last two 
times in the simulation. The optimum values for system with
$n=4$ components are $\delta= 0.435$ with a Pearson coefficient
of $ -.999998$. The other two parameters are $\ln A= 13.21$ and $b=8.19$
This result is presented in the Fig. 2(b).

We turn now to the description  of the case $n=5$, following a similar 
analysis to the $n=4$ model. Fig. 3(a) shows the correlation function
as a function of distance  for different times. In this model the collapse
is also achieved using the characteristic length $L(t)= t^{1/2}$, as can be
observed in  Fig.3(b). Therefore, both models are consistent
with dynamical exponent $z=2$. The corresponding
scaling plot for the structure factor is shown in Fig. 4(a).
A more important effect is observed in the  
structure factor tail, because in this case it  
has also a stretched exponential but with an apparently larger exponent. 
Following an analysis similar to that used for $n=4$, we find that 
the value of the best fit value of the exponent is $ \delta= 0.613$, 
and the corresponding PCC in the regression is $-0.999998$.
The other two parameters are $\ln A= 7.57$ and $b= 4.39$. 
In  Fig. 4(b) we plot  $\ln g$ against  
$q^\delta$ and the linear behaviour is clearly seen.

Comparison between the real-space correlation functions
of the $n=4$ and $n=5$ models shows that the scaling functions are very 
similar; the main difference is that the scaling function decreases 
slightly more slowly for $n=5$ than $n=4$. 
This is reflected in the parameters of the fitting
function for the  small-$x$  range: the amplitudes $\alpha$ and $\beta$
tend to decrease as $n$ increases (table 1). 

We turn to the discussion of the simulation results for one-dimensional 
systems. We shall describe the relevant behavior in Fourier space. 
Real-space data has been presented in \cite{NEW}.  
Our results show that in one-dimensional systems  the asymptotic behavior 
of the structure factor also has a stretched exponential form, but the 
fitted exponents $\delta$ are larger than those of the 
corresponding two-dimensional models, and close to unity for $n=4$ and 5. 

In Fig.\ 5(a), we present the simulation results for the 
scaling function $g(q)$ of the structure factor for the one-dimensional 
Heisenberg Model ($n=3$). The continuous curve is the result of the 
analytical approach described in section IV. 
The analysis of the asymptotic behaviour gives an exponent $\delta= 0.79$ 
for the stretched exponential. Fig.\ 5(b) shows the
curve of $\ln[g(q)]$ versus $q^\delta$, where the linear behaviour is
clearly observed. Similarly, we present the corresponding plots for the   
$n=4$ model in Fig.\ 6, where the measured exponent is now $\delta = .98$, 
while for $n=5$ we obtain $\delta = 1.02$ as is shown in Fig.\ 7. 
The values of $\delta$ for the last two models are so  close that in 
practice it is difficult to distinguish between them. They are also 
close to the value unity obtained from the approximate large-$n$ equation 
discussed in the following section. 
    
It is clear from the  results for the one-dimensional
models  that the scaling function in real space is not a gaussian 
function,  despite the good real-space fits to this form obtained 
in \cite{NEW}. Moreover, the (effective) exponents $\delta$ for the $n=4,5$ 
models are bigger than for the corresponding models in two dimensions. 
Therefore, we have evidence that the exponent  $\delta$ 
increases with $n$, while it seems to decrease with $d$.
Note that the gaussian result obtained for $n=\infty$ corresponds 
to $\delta=2$, so the results presented here for the structure-factor tail 
are actually quite far from that limit. The analytical treatment presented 
in the next section gives some indication of why this might be expected. 
In particular, it suggests that the structure factor is dominated by a 
simple exponential for $q \to \infty$ at fixed large $n$, while the familiar 
gaussian form is recovered as $n \to \infty$ at fixed $q$.
  
We conclude this section by discussing briefly some alternative fitting 
forms for the structure factor tail. First, however, we note that the 
stretched exponential form (\ref{se}) describes the tail well over at 
least 10 decades of $S(k)$ in all cases. Of course, this represents 
a much smaller dynamic range (1 to $1\frac{1}{2}$ decades) in the 
scaling variable $q=kL(t)$. Motivated by the analytical result [equation 
(\ref{g}) below] of the approximate large-$n$ theory, other fitting forms 
were tried for $d=2$ (the agreement with the large-$n$ theory already 
being good for $n =4$ and 5 in $d=1$). A direct fit of (\ref{g}) does not 
work well for $d=2$. Allowing for a general power-law prefactor, 
$g(q) = Aq^{-x}\exp(-bq)$, gives a reasonable fit, but with very large 
values for $x$ --- 5.6 for $n=4$, and 6.7 for $n=5$. Fixing $x=1/2$, but 
allowing for a general stretched exponent $\delta$ again gives a reasonable 
fit (with $\delta \simeq 0.68$ for $n=4$ and 0.70 for $n=5$), but over a 
significantly reduced range of $q$. For these reasons we prefer the 
unmodified stretched exponential (\ref{se}) as giving the simplest and most 
convincing description of the large-$q$ data, at least for these small values 
of $n$ in $d=2$. Of course, it is quite possible that the form (\ref{g}) 
will fit the data well at larger values of $n$. 

\section{ANALYTICAL TREATMENT} 
In an attempt to gain some analytical insight into the structure factor 
asymptotics, we start from an approximate equation of motion for the 
pair correlation function derived using the gaussian auxiliary field 
approach pioneered by Mazenko \cite{Mazenko}. 
We then make, for reasons that will become clear, the 
further simplification of retaining only the leading nonlinearity as 
$n \to \infty$. The resulting equation is then finally used to extract 
the asymptotics of $g(q)$.

The GAF method for vector fields has been discussed in some detail 
elsewhere. We refer the reader to the original papers \cite{GAF} and a 
recent review \cite{REV} for a full exposition. The essence of the method 
is a mapping from the original field variable $\vec{\phi}$ to an `auxiliary 
field' $\vec{m}$. The function $\vec{\phi}(\vec{m})$ satisfies the equation 
$(1/2)\sum_{i=1}^n \partial^2 \vec{\phi}/\partial_{m_i}^2 
= \partial V/\partial\vec{\phi}$, where $V(\vec{\phi})$ is the potential in the 
Ginzburg-Landau function (4). With the boundary conditions $\vec{\phi}(0)=0$, 
and $\vec{\phi}(\vec{m}) \to \vec{m}/|\vec{m}|$ for $|\vec{m}| \to \infty$, 
this equation for $\vec{\phi}(\vec{m})$ represents the equilibrium profile 
function for a spherically symmetric topological defect, with 
$|\vec{m}|$ representing distance from the defect. 

The (uncontrolled) approximation that $\vec{m}$ is a gaussian field, 
and the imposition of the scaling form (1), leads eventually to the 
self-consistent equation \cite{REV,GAF}
\begin{equation}
f'' + \left(\frac{d-1}{4} + \frac{x}{4}\right)f' 
+ \frac{\lambda}{2}\, \gamma \frac{dC}{d\gamma} = 0
\label{SC1}
\end{equation}
for the scaling function $f(x)$, where primes indicate derivatives. 
In (\ref{SC1}) $\gamma$ is the normalized correlator of the auxiliary field, 
$\gamma = \langle \vec{m}(1)\cdot\vec{m}(2) \rangle
/[\langle \vec{m}^2(1)\rangle \langle \vec{m}^2(2) \rangle]^{1/2}$, 
where `1' and `2' represent the space-time points ${\bf x}_1,t$ and 
${\bf x}_2,t$, and the function $C(\gamma)$ is given by
\begin{equation}
C(\gamma) = \frac{n\gamma}{2\pi}\,
\left[B\left(\frac{n+1}{2},\frac{1}{2}\right)\right]^2\,
F\left(\frac{1}{2},\frac{1}{2};\frac{n+2}{2};\gamma^2\right)\ ,
\label{BPT}
\end{equation}
where $B(x,y)$ is the beta function, and $F(a,b;c;z)$ the hypergeometric 
function. The constant $\lambda$ in (\ref{SC1}) has to be adjusted so 
that $f(x)$ vanishes sufficiently fast at infinity \cite{GAF}. 

As should be clear from the above discussion, (\ref{SC1}) only really makes 
sense for $n \le d$, based as it is on the presence of singular topological 
objects whose positions are defined by the zeros of the field $\vec{\phi}$ or, 
equivalently, by the zeros of $\vec{m}$. Indeed, the function $C(\gamma)$ 
has inbuilt structure that generates the Porod tail associated with 
such defects. Specifically, in the short-distance limit, where $\gamma \to 1$, 
the hypergeometric function in (\ref{BPT}) has a singular contribution of 
order $(1-\gamma^2)^{n/2}$ (with a logarithmic correction for even $n$). 
Since $1-\gamma^2 \sim x^2$ for small scaling variable $x$, this singular 
term is of order $x^n$ (again, with a logarithm for even $n$), leading to the 
power-law tail $g(q) \sim q^{-(d+n)}$ in Fourier space. Within the GAF 
approach, this tail is obtained for {\em all} $n$ and $d$. For $n>d+1$, 
however, neither singular topological objects nor nonsingular topological 
textures exist, so the GAF result is qualitatively incorrect. Indeed, 
this is to be expected since the GAF approach is specifically designed 
to build in the defect structure. 

So what should one do when there are no defects? We have seen that the 
usual GAF approach always give a Porod tail, for any $n$ and $d$: this is 
unphysical for $n>d+1$, since the tail is a consequence of the presence of 
topological defects. One way around this impasse is to artificially 
approximate the full GAF equation (\ref{SC1}) by the form valid for 
$n \to \infty$. In this limit $\gamma dC/d\gamma = f +f^3/n +O(1/n^2)$, 
and (\ref{SC1}) becomes, correct to $O(1/n)$,  
\begin{equation}
f'' + \left(\frac{d-1}{4} + \frac{x}{4}\right)f' 
+ \frac{\lambda}{2}\,\left(f + \frac{1}{n}f^3\right) = 0\ .
\label{SC2}
\end{equation}
This step, admittedly ad-hoc, has the desired effect of eliminating 
the unwanted (for $n>d+1$) short-distance singularity in $f(x)$. 
Eq.\ (\ref{SC2}) is the nonconserved version of the equation introduced 
by BH to study the crossover from multiscaling to simple scaling 
in the asymptotic dynamics of a {\em conserved} vector field at large but 
fixed $n$ \cite{BH92}. Both conserved and nonconserved versions have 
recently been studied numerically \cite{Cast}. 

To extract analytically the large-$q$ behavior, we perform a 
($d$-dimensional) Fourier transform of (\ref{SC2}). The resulting 
equation for $g(q) \equiv \int d^dx\,f(x)\exp(i{\bf q}\cdot{\bf x})$ is 
\begin{equation}
\left(\frac{d}{4}+q^2\right)g(q) + \frac{q}{4} g'(q) = 
\frac{\lambda}{2}\,\left(g(q) + B(q)\right)\ ,
\label{SC3}
\end{equation}
where
\begin{equation}
B(q) = \frac{1}{n} \int d^dx\,f^3(x)\exp(i{\bf q}\cdot{\bf x})\ .
\label{B}
\end{equation}
If we assume an asymptotic form $g(q) \sim q^\nu\exp(-bq^\delta)$, with 
$\delta<2$, then (\ref{SC3}) gives
\begin{equation}
B(q) \to \frac{2q^2}{\lambda} g(q)\ ,\ \ \ \ \ q \to \infty\ .
\label{largeq}
\end{equation}

In the Appendix, we show that consistency with (\ref{largeq}) requires 
$\delta=1$ and $\nu = (1-d)/2$, i.e. 
\begin{equation}
g(q) \to Aq^{(1-d)/2}\exp(-bq)\ ,\ \ \ \ \ q \to \infty\ .
\label{g}
\end{equation}  
In real-space this implies that the function $f(z)$ has simple poles 
in the complex $z$ plane at $z=\pm ib$. The value of $b$ is 
not determined by this argument; instead one can derive (see Appendix) 
the relationship 
\begin{equation}
A^2 = (16\pi^2 n/\lambda)\,(2\pi b)^{d-1}
\label{A}
\end{equation}
between $b$ and the prefactor $A$ in the asymptotic form (\ref{g}). 
The existence of these simple poles in real space also follows directly from 
the real-space equation (\ref{SC2}). If one assumes a singularity of the 
form $(z-z_0)^{-\gamma}$, with $\gamma>0$,  then balancing the dominant terms 
$f''$ and $\lambda f^3/2n$ gives immediately $\gamma=1$, i.e.\ a simple pole. 
The position $z_0$ is not determined, but the residue $C$ of the pole is 
given by $C=\mp i(4n/\lambda)^{1/2}$, where the two values correspond to the 
poles $z_0 = \pm ib$. Using this result for $C$, one can readily recover 
(\ref{A}) by contour methods, e.g.\ for $d=1$ one has
\begin{eqnarray}
g(q) & = & \int_{-\infty}^{+\infty} dx\,f(x)\exp(iqx) \nonumber \\
     & \to & 2\pi(4n/\lambda)^{1/2}\exp(-bq)\ ,\ \ \ \ q \to \infty\ ,
\end{eqnarray}
where the second line, equivalent to (\ref{A}) for $d=1$, was obtained by 
closing the contour in the upper half-plane. 

The approach outlined above gives the relation (\ref{A}) between $A$ and 
$b$, but does not determine $b$ explicitly. We now give a heuristic 
argument that $b \sim (\ln n)^{1/2}$ for large $n$. First we make an 
observation concerning the value of $\lambda$. Equation (\ref{SC3}) with 
$q=0$ gives 
\begin{eqnarray}
\lambda & = & \frac{d}{2}\,\frac{g(0)}{g(0)+B(0)} \nonumber \\
        & = & \frac{d}{2}\,\left[1 + \frac{1}{n}\int d^dx\,
               f^3(x)/ \int d^dx\,f(x)\right]^{-1}\ .
\end{eqnarray}
In particular, $\lambda = d/2$ for $n=\infty$. For $n=\infty$, therefore, 
(\ref{SC3}) becomes $q^2g(q)+(q/4)g'(q)=0$, with solution 
$g(q) = (8\pi)^{d/2}\exp(-2q^2)$ [the prefactor being fixed by the 
condition $f(0)=1$]. For $n$ large but finite, on the other hand, we have 
seen that the asymptotic form is $g(q) \sim \exp(-bq)$. The crossover 
between these two forms presumably occurs at some $q=q^*(n)$, with 
$g(q) \sim \exp(-2q^2)$ for $1 \ll q \ll q^*$, and $g(q) \sim \exp(-bq)$ 
for $q \gg q^*$. Matching these two forms at $q=q^*$ gives $q^* \sim b$. 
Next we evaluate $B(q)$ in the region $q \ll q^*$. 
Here $q(q) \simeq (8\pi)^{d/2}\exp(-2q^2)$, so $f(x) \simeq \exp(-x^2/8)$, 
giving $B(q) \sim (1/n)\exp(-2q^2/3)$. However, this decays more slowly 
with $q$ than the other terms in (\ref{SC3}), which fall off as 
$\exp(-2q^2)$. So the term involving $B(q)$ (evaluated for $q \ll q^*$) 
becomes comparable with the other terms when $(1/n)\exp(-2q^2/3) 
\sim \exp(-2q^2)$, i.e.\ when $q \sim (\ln n)^{1/2}$. This suggests 
$q^* \sim (\ln n)^{1/2}$, and therefore $b \sim (\ln n)^{1/2}$. 
The numerical data of Castellano and Zannetti \cite{Cast} certainly show 
that $b$ increases extremely slowly with $n$. 

To compare this approximate theory with our simulation data, we have solved 
(\ref{SC1}) numerically for $d=1$, $n=3,4,5$, using the procedure described 
in \cite{GAF}. The Fourier transform was then taken numerically, and the 
`best fit by eye' to the structure-factor data was obtained by adjusting 
the timescale in the theoretical curves, giving the results shown by the
continuous curves in Figures 5(a), 6(a) and 7(a). The corresponding 
log-linear plots, which reveal the large-$q$ behaviour more clearly, 
are shown in Figures 8(a)-(c). As might be expected, equation 
(\ref{SC1}) [or its Fourier transform (\ref{SC3})] does not describe the 
data quantitatively over the whole range of $q=kL(t)$, but it does give a 
qualitatively correct description. There is an early parabolic region, 
corresponding to a gaussian form for $g(q)$, which then gives way to a slower 
decay that, at least for $n=4$ and 5, is consistent with the simple 
exponential form predicted by (\ref{SC3}) but with a different coefficient 
$b$ in the exponent. Given that the theory is, at best, a large-$n$ theory 
we regard these results as encouraging.
The $n=3$ data, however, and the $d=2$ data, do not 
seem to fit a simple exponential, at least for the range of $q$ that we 
have been able to explore. (This is of course implicit in the values 
$\delta<1$ obtained for these systems from Figures 2,4 and 5.) \ It may 
well be that considerably larger values of $n$ are needed in $d=2$ than 
in $d=1$ for the large-$n$ asymptotics to become apparent.  

The above derivation of an exponential tail was specific to nonconserved 
fields. What can we say for conserved fields? The fundamental equation 
of motion for this case is obtained from the TDGL equation (\ref{TDGL}) 
by the replacement $\Gamma \to -\Gamma \nabla^2$. Applying the GAF method 
to this equation, imposing the scaling form (\ref{scaling:real}) [but 
with $L(t) = t^{1/4}$ for conserved fields], and taking the Fourier 
transform, leads to \cite{BH92}
\begin{equation}
\left(\frac{d}{8}+q^4\right)g(q) + \frac{q}{8} g'(q) = 
\frac{\lambda}{2}q^2\,\left(g(q) + B(q)\right)\ ,
\label{CONS}
\end{equation}
instead of (\ref{SC3}). [The definition (\ref{B}) of $B(q)$ differs by a 
constant from that used in \cite{BH92}, where $\lambda$ was written as 
$2q_m^2$, $q_m$ being the position of the maximum of $g(q)$ for large $n$.] \  
Assuming the asymptotic form $q(q) \sim q^\nu \exp(-bq^\delta)$ 
for $q \to \infty$, (\ref{CONS}) gives (\ref{largeq}) once more, provided 
$\delta <4$. Then our previous arguments apply,  and the asymptotic form 
(\ref{g}), with $A$ and $b$ related by (\ref{A}), are recovered. This 
approach therefore predicts that the structure factors for conserved and 
nonconserved systems will have the {\em same} asymptotic forms, at least 
within the context of the BH truncation. The same conclusion was drawn 
from recent numerical solutions of the BH equation \cite{Cast}. 

\section{CONCLUSION}
In summary, we have studied the dynamics of phase ordering for models 
without topological defects in one and two dimensions. We find that 
scaling is achieved with the growth law $L(t)= t^{1/2}$.
The tail in the structure factor is well fitted by a stretched
exponential form. For the two-dimensional systems, table 1 summarizes 
the relevant parameters describing the fits in real and Fourier 
space. In contrast to systems with singular defects ($n\le d$), where the 
generalized Porod form $g(q) \sim q^{-(d+n)}$ for the structure factor tail 
is a consequence of the defect structure, and is independent of the presence 
or absence of conservation laws, in systems without defects the functional 
form does, apparently, differ for conserved and nonconserved systems. 
We have shown, for example, that for the particular case of the $n=4$ model 
in two dimensions the tail is well described, over the range of $q$ 
accessible to us, by a stretched exponential with exponent $\delta=.435$, 
differing from the result for the same model with conservation studied 
by RC \cite{RC}, who found $\delta \simeq 1.7$. 

Within the `toy' equation of Bray and Humayun \cite{BH92}, however, we have 
shown that the true asymptotics are the {\em same} for conserved and 
nonconserved dynamics. Of course, the BH equation is at best a large-$n$ 
theory, and the numerical results for nonconserved and conserved 
dynamics may converge as $n$ is increased. 
A related question is whether the exponents $\delta$ measured here and in 
\cite{RC} are genuine asymptotic exponents, or effective exponents whose 
values will change as the range of $q$ over which the fit is made is moved 
to larger $q$. More extensive simulations may cast some light on this issue. 
The `universal' (independent of conservation laws) Porod tail behavior 
obtained for $n \le d$ is  geometrical in origin, being a consequence of 
the field structure induced by singular topological defects \cite{BH}. 
As yet, however, we have no corresponding physical picture in the absence 
of topological defects.  

It is interesting that, within the simple model of equation (\ref{SC1}), 
the exponent $\delta$ jumps discontinuously from $\delta=2$ at $n=\infty$ 
to $\delta=1$ for $n$ large but finite. More precisely, one can say that 
$\delta=2$ corresponds to the limit $n \to \infty$ at fixed, large $q$, 
while $\delta = 1$ corresponds to $q \to \infty$ at fixed, large $n$. 
We have argued that the crossover between these limiting forms for 
fixed, large $n$ occurs at $q \sim (\ln n)^{1/2}$. 
This change of behavior depending on the order of the limits is 
reminiscent of the result obtained from the conserved version of 
(\ref{SC1}), where a novel `multiscaling' behavior is obtained for 
$n \to \infty$ at fixed, large $t$ \cite{CZ}, while simple scaling is 
recovered for $t \to \infty$ at fixed, large $n$ \cite{BH92}. For the 
nonconserved case, one always has simple scaling. For both conserved 
and nonconserved fields, however, the asymptotics of $g(q)$ are 
sensitive to whether $n$ is large or truly infinite. This rules out, 
for example, exploring the asymptotics by expanding around the large-$n$ 
solution in powers of $1/n$.

\acknowledgments

We thank Sanjay Puri for stimulating discussions during the early 
stages of this work. F. Rojas thanks CONACYT (Mexico) for financial 
support. This work was supported by EPSRC (UK) grant GR/J24782.

\section{APPENDIX}
In this Appendix we use (\ref{largeq}) to derive the asymptotic form 
(\ref{g}) for $g(q)$. From the definition (\ref{B}) we have 
\begin{equation}
B(q) = \frac{1}{n} \int_{\bf p}\int_{\bf k} g({\bf p})g({\bf k})
g({\bf q}-{\bf p}-{\bf k})\ ,
\end{equation}
where $\int_{\bf p} \equiv \int d^dp/(2\pi)^d$. Inserting the asymptotic 
form
\begin{equation}
g(q) \to Aq^\nu\exp(-bq^\delta)\ ,
\end{equation} 
gives 
\begin{equation}
B(q) \to \frac{A^3}{n}\int_{\bf p}\int_{\bf k}F({\bf p},{\bf k},{\bf q})\,
\exp[-bE({\bf p},{\bf k},{\bf q})]\ ,
\end{equation}
where 
\begin{eqnarray}
F({\bf p},{\bf k},{\bf q}) & = & 
|{\bf p}|^\nu|{\bf k}|^\nu|{\bf q}-{\bf p}-{\bf k}|^\nu\ \nonumber \\
E({\bf p},{\bf k},{\bf q}) & = &
|{\bf p}|^\delta + |{\bf k}|^\delta + 
|{\bf q}-{\bf p}-{\bf k}|^\delta\ .
\end{eqnarray}
We now scale out the $q$-dependence through the changes of variable 
${\bf p}=q{\bf u}$, ${\bf k}=q{\bf v}$, ${\bf q}=q{\bf e}$, where 
${\bf e}$ is a unit vector. Then 
\begin{equation}
B(q) = \frac{A^3}{n}\,q^{2d+3\nu}\int_{\bf u}\int_{\bf v}
F({\bf u},{\bf v},{\bf e})\, 
\exp[-bqE({\bf u},{\bf v},{\bf e})]\ .
\end{equation}

For $q \to \infty$, we can attempt to evaluate the ${\bf u}$ and ${\bf v}$ 
integrals using the method of steepest descents. This requires minimizing 
the function $E({\bf u},{\bf v},{\bf e})$. The points requiring 
consideration are the symmetry point, ${\bf u}={\bf v}={\bf e}/3$, and 
the points ${\bf u}={\bf 0}={\bf v}$ and two similar points obtained by 
permuting ${\bf u}$, ${\bf v}$ and ${\bf e}-{\bf u}-{\bf v}$. The 
corresponding values of $E$ are $E({\bf e}/3,{\bf e}/3,{\bf e}/3) 
= 3^{1-\delta}$, and $E({\bf 0},{\bf 0},{\bf e}) = 
E({\bf 0},{\bf e},{\bf 0}) = E({\bf e},{\bf 0},{\bf 0})=1$. 
Thus for $\delta >1$, the symmetry point minimizes $E$, giving 
$B(q) \sim \exp(-3^{1-\delta}bq)$. But this form violates the asymptotic 
relation (\ref{largeq}), according to which $B(q)$ and $g(q)$ must decay 
with the {\em same} exponential factor, so $\delta>1$ is ruled out. 

For $\delta<1$, the smallest $E$ is unity, obtained when two of ${\bf u}$, 
${\bf v}$, and ${\bf e}-{\bf u}-{\bf v}$ vanish. So this case is apparently 
consistent with (\ref{largeq}). However, the integral is now dominated by 
points where two of the momenta ${\bf p}$, ${\bf k}$, and ${\bf q}-{\bf p} 
-{\bf k}$ vanish. This invalidates the use of the asymptotic form for $g(q)$ 
in the evaluation of $B(q)$, so the derivation of a stretched exponential 
form is not internally consistent for $\delta<1$.

This leaves $\delta=1$.  For this case all points of the form 
${\bf u}=\alpha {\bf e}$, ${\bf v}=\beta {\bf e}$, with $0\leq\alpha\leq 1$ 
and $0\leq\beta\leq 1-\alpha$, give $E=1$, so one has to integrate over all
such points. Writing ${\bf u}=\alpha{\bf e} + {\bf u}_\perp$, 
${\bf v} = \beta{\bf e} + {\bf v}_\perp$, expanding $E$ to quadratic 
order in ${\bf u}_\perp$, ${\bf v}_\perp$, and carrying out the integrals 
over ${\bf u}_\perp$, ${\bf v}_\perp$, gives after some algebra
\begin{eqnarray}
B(q) & = & (A^3/4\pi^2n)\,q^{d+1+3\nu}\exp(-bq)\,I(d,\nu)/(2\pi b)^{d-1} 
\nonumber \\
I(d,\nu) & = & \int_0^1 d\alpha\int_0^{1-\alpha}d\beta\,
[\alpha\beta(1-\alpha-\beta)]^{\nu+(d-1)/2}\ .
\label{B1}
\end{eqnarray} 
But (\ref{largeq}) implies, asymptotically,
\begin{equation}
B(q) = (2A/\lambda)\,q^{2+\nu}\,\exp(-bq)\ .
\label{B2}
\end{equation}
Comparing (\ref{B1}) and (\ref{B2}) gives $\nu = (1-d)/2$ and Eq.\ (\ref{A}) 
for the amplitude $A$.

\newpage
\begin{center}
{FIGURES}
\end{center}

FIG.\ 1. (a) Real-space correlation function $C(r)$ for a 2D system with 
$n=4$ components as a function of distance $r$ for several times $t$. 
The data were obtained from  lattices of size $256\,\times\,256$,
averaged over 20 different initial conditions. (b) Demonstration of 
dynamic scaling, $C(r,t) = f(r/L(t))$, with  $L(t)= t^{1/2}$.  
          
\smallskip

FIG.\ 2. (a) Scaling structure factor $g(q) = [L(t)]^{-2}\,S(k,t)$ as a 
function of scaled momentum $q=kL(t)$ for a  2D system with
$n=4$ components for lattices of size 256$\,\times\,$256 (averaged over 
20 different initial conditions). 
(b) Demonstration of the stretched exponential behaviour, 
plotting $\ln g(q)$ against $q^{\delta}$ with $\delta= 0.435$. The  
line is included as a guide to the eye. 

\smallskip

FIG.\ 3. Same as Fig.\ 1 but for the $n=5$ model.

\smallskip

FIG.\ 4. (a) Same as Fig.\ 2 but for the $n=5$ model.
(b) The power $\delta =0.613 $ was found to give the best linear 
relation between $\ln g(q)$ and $q^\delta$. 

\smallskip

FIG.\ 5. (a) Scaling structure factor $g(q) = [L(t)]^{-1}\,S(k,t)$ as a 
function of scaled momentum $q=kL(t)$ for a 1D Heisenberg  model
[$O(3)$ model] for lattices of size 16384 (averaged over 100 different
initial conditions). Continuous curve: result of the approximate 
analytical treatment described in the text. 
(b) Demonstration of the stretched exponential
behaviour, plotting $\ln g(q)$ against $q^{\delta}$ with $\delta= 0.79$. The  
line is a guide to the eye. 
  
\smallskip

FIG.\ 6. (a) Same as Fig.\ 5, but for the $n=4$ model.
(b) The power  $\delta =0.98 $ is found for this model
to give the best linear relation between $\ln g(q)$ and $q^\delta$. 

\smallskip

FIG.\ 7. (a) Same as Fig.\ 5 but for the $n=5$ model.
(b) Same as Fig.\ 5(b), but with $\delta =1.02 $. 

\smallskip

FIG.\ 8. Log-linear plot of the scaled structure factor against scaled 
momentum for the 1D systems: (a) $n=3$ (b) $n=4$ (c) $n=5$. In each case 
the continuous curve is the result of the approximate theory of section IV.

\begin{table}
\caption{ Parameters determined from fitting of the simulation results 
for the two-dimensional
systems : the  small-x solution ($x<0.5$) in the scaling function with the 
form $f(x) = 1-\alpha x^2 + \beta x^4$ and
the  asymptotic analysis of scaling structure factor in term of the
stretched  exponential
$g(q)\sim \exp(-bq^\delta)$}

\begin{tabular}{ddddd} 
\multicolumn{1}{c}{} 
&\multicolumn{2}{ c}{small-x}&\multicolumn{2}{c}{Tail}\\  
$n$  &  $\alpha$  & $\beta$        & $\delta$ &   $b$ \\             
\tableline 
 4   &  1.5326    & 1.3916         & 0.435    &  8.19       \\
 5   &  1.3040    & 0.8417         & 0.613    &  4.39       \\ 
 \end{tabular}
\label{tab:ajus}
\end{table}

\newpage

\begin{references}
\bibitem{REV} 
For a recent review see A. J. Bray, Adv.\ Phys.\ {\bf 43}, 357 (1994). 

\bibitem{MG} 
M. Mondello and N. Goldenfeld, Phys. Rev. A {\bf 42}, 5865 (1990); 
{\bf 45}, 657 (1992); Phys.\ Rev.\ E {\bf 47}, 2384 (1993); 
R. E. Blundell and A. J. Bray, Phys.\ Rev.\ E {\bf 49}, 4925 (1994).

\bibitem{BH90} 
A. J. Bray and K. Humayun, J.\ Phys.\ A {\bf 23}, 5897 (1990).

\bibitem{VEC} M. Siegert and M. Rao, Phys.\ Rev.\ Lett.\ {\bf 70}, 
1956 (1993); S. Puri, A. J. Bray and F. Rojas, Phys.\ Rev.\ E {\bf 52}, 
4699 (1995). 

\bibitem{NEW} 
T.J. Newman, A.J. Bray, and M. A. Moore, Phys.\ Rev.\ B {\bf 42}, 4514 (1990). 
    
\bibitem{Porod} 
A. J. Bray and S. Puri, Phys.\ Rev.\ Lett.\ {\bf 67}, 2670 (1991); 
H. Toyoki, Phys.\ Rev.\ B {\bf 45},1965 (1992).

\bibitem{BH} 
A. J. Bray and K. Humayun, Phys.\ Rev.\ E {\bf 47}, R9 (1993).

\bibitem{OXY} 
A. D. Rutenberg and A. J. Bray , Phys.\ Rev.\ Lett.\ {\bf 74}, 3836 (1995). 

\bibitem{TOT} 
A. D. Rutenberg, Phys.\ Rev.\ E {\bf 51}, R2715 (1995); see also 
M. Zapotocky and W. Zakrzewski, Phys.\ Rev.\ E {\bf 51}, R5189 (1995). 

\bibitem{RC} 
M. Rao and A. Chakrabarti Phys.\ Rev.\ E {\bf 49}, 3727 (1994).

\bibitem{CZ} 
A. Coniglio and M. Zannetti, Europhys.\ Lett.\ {\bf 10}, 575 (1989). 

\bibitem{BH92} 
A. J. Bray and K. Humayun, Phys.\ Rev.\ Lett.\ {\bf 68}, 1559 (1992). 

\bibitem{Cast} 
C. Castellano and M. Zannetti, preprint. 

\bibitem{GAF} 
F. Lui and G. F. Mazenko, Phys.\ Rev.\ B {\bf 45}, 6984 (1992); 
A. J. Bray and K. Humayun, J.\ Phys.\ A {\bf 25}, 2191 (1992).

\bibitem{TU} 
H. Toyoki, Mod.\ Phys.\ Lett.\ B {\bf 7}, 397 (1993), and in {\it Formation, 
Dynamics and Statistics of Patterns} vol.\ 2, edited by K. Kawasaki, 
M. Suzuki and A. Onuki (World Scientific, Singapore, 1994).

\bibitem{Nematics} 
A. P. Y. Wong, O. Wiltzius and B. Yurke, Phys.\ Rev.\ Lett.\ {\bf 68}, 
3583 (1992); 
N. Mason, A. N. Pargellis and B. Yurke, Phys.\ Rev.\ Lett.\ {\bf 70}, 
190 (1993); 
R. E. Blundell and A. J. Bray, Phys.\ Rev.\ A {\bf 46}, R6154 (1992).

\bibitem{AR}
A. J. Bray  and A. D. Rutenberg, Phys.\ Rev.\ E {\bf 49}, 27 (1994); 
Phys.\ Rev.\ E {\bf 51}, 5499 (1995). 

\bibitem{BREN}
A. J. Bray, Phys.\ Rev.\ Lett.\ {\bf 62}, 2841 (1989).

\bibitem{OP1}
Y. Oono and S. Puri, Phys.\ Rev.\ Lett.\ {\bf 58}, 836 (1987); 
Phys.\ Rev.\ A {\bf 38}, 434 (1988); 
S. Puri and Y. Oono, Phys.\ Rev.\ A {\bf 38}, 1542 (1988).

\bibitem{Mazenko}
G. F. Mazenko, Phys.\ Rev.\ Lett.\ {\bf 63}, 1605 (1989); 
Phys.\ Rev.\ B {\bf 42}, 4487 (1990); {\bf 43}, 5747 (1990). 

\end{references}
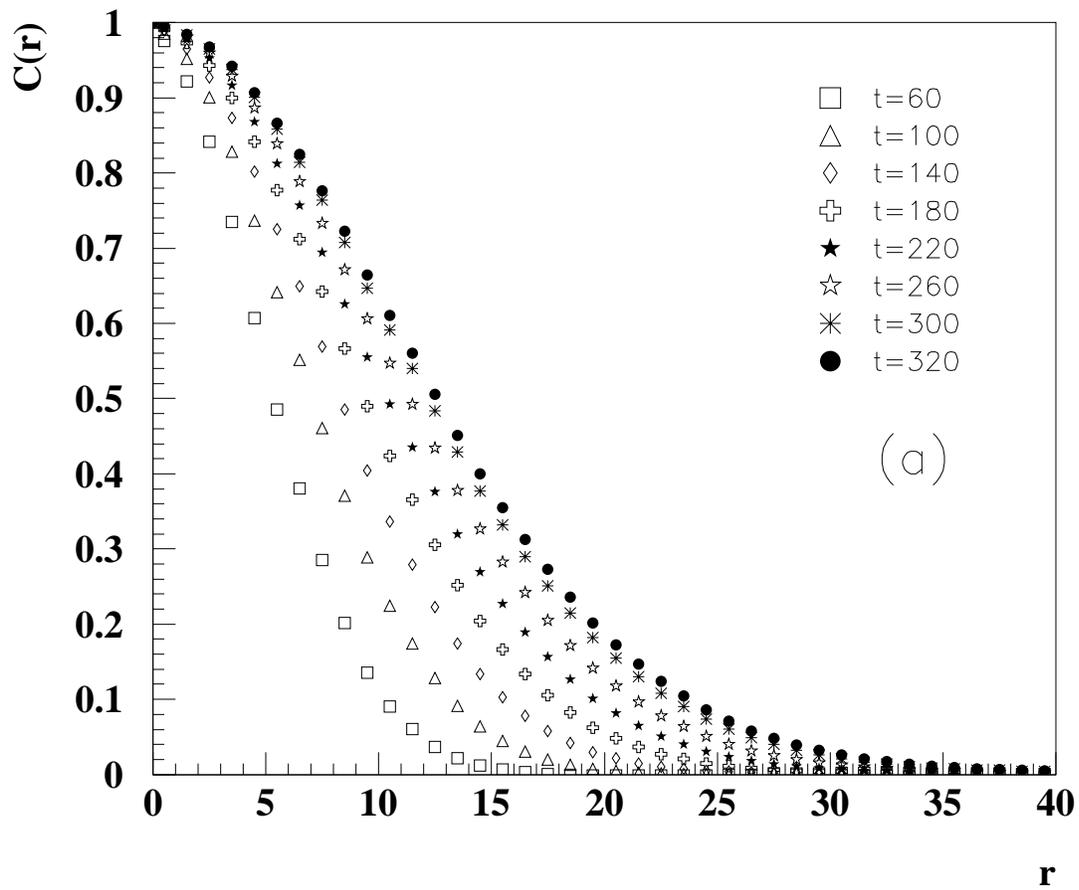
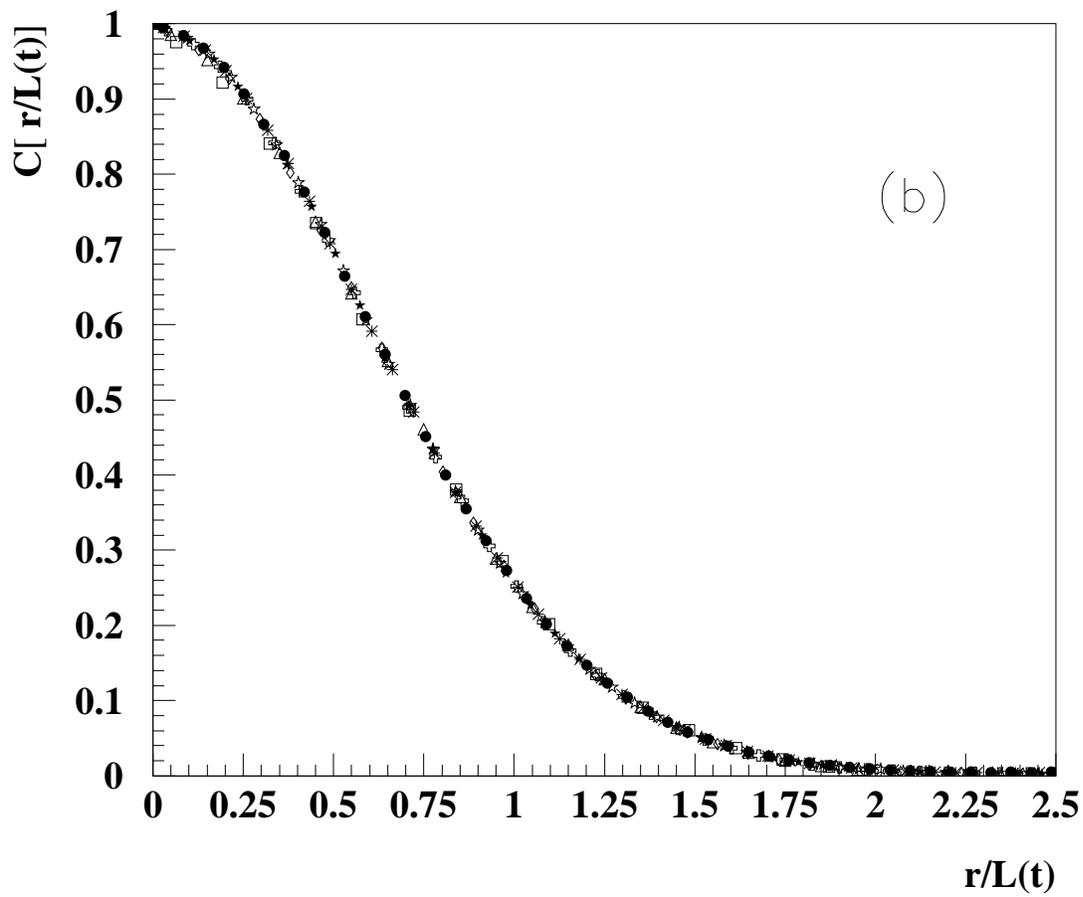
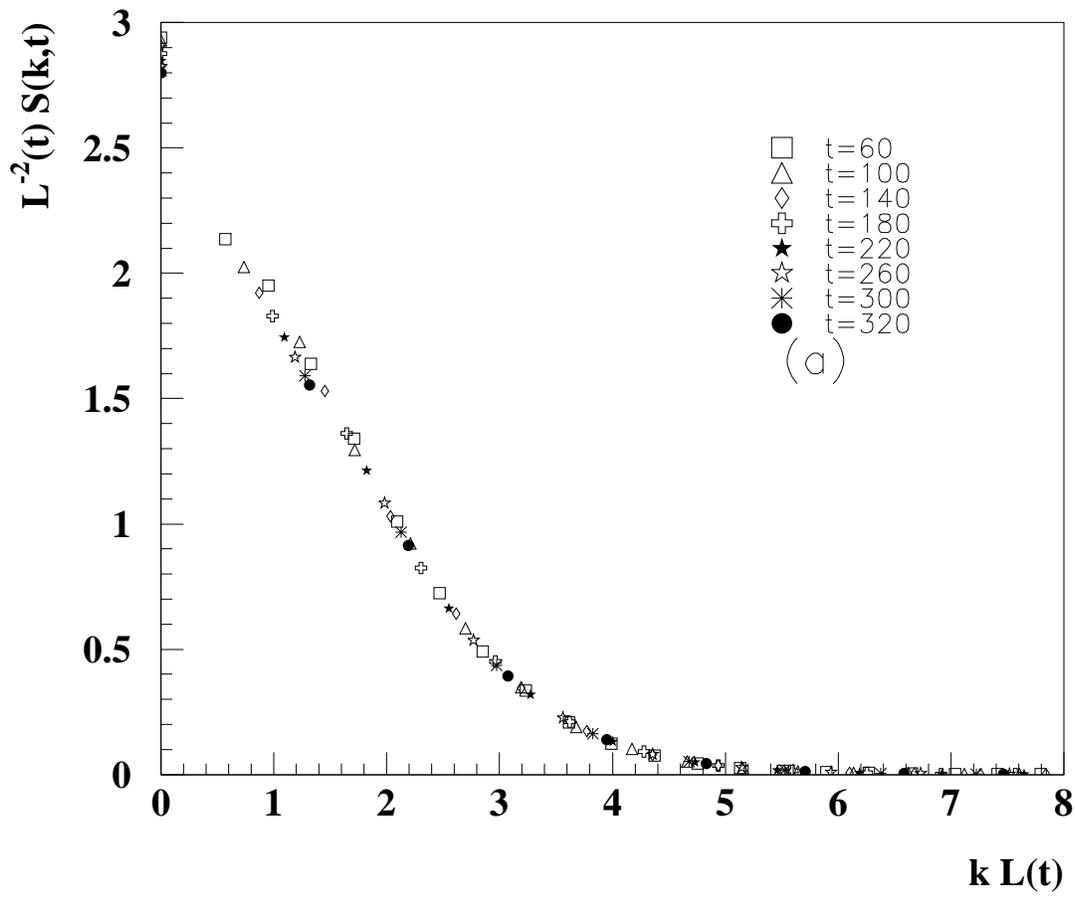
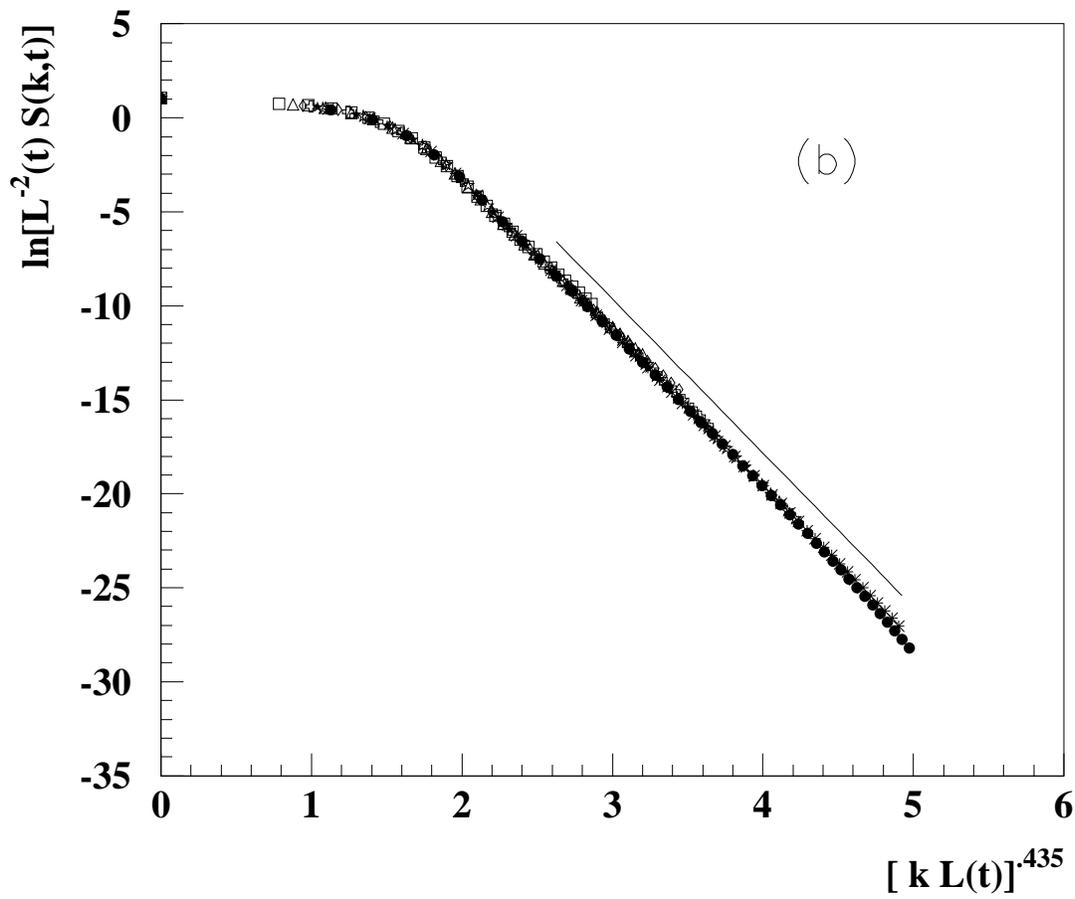
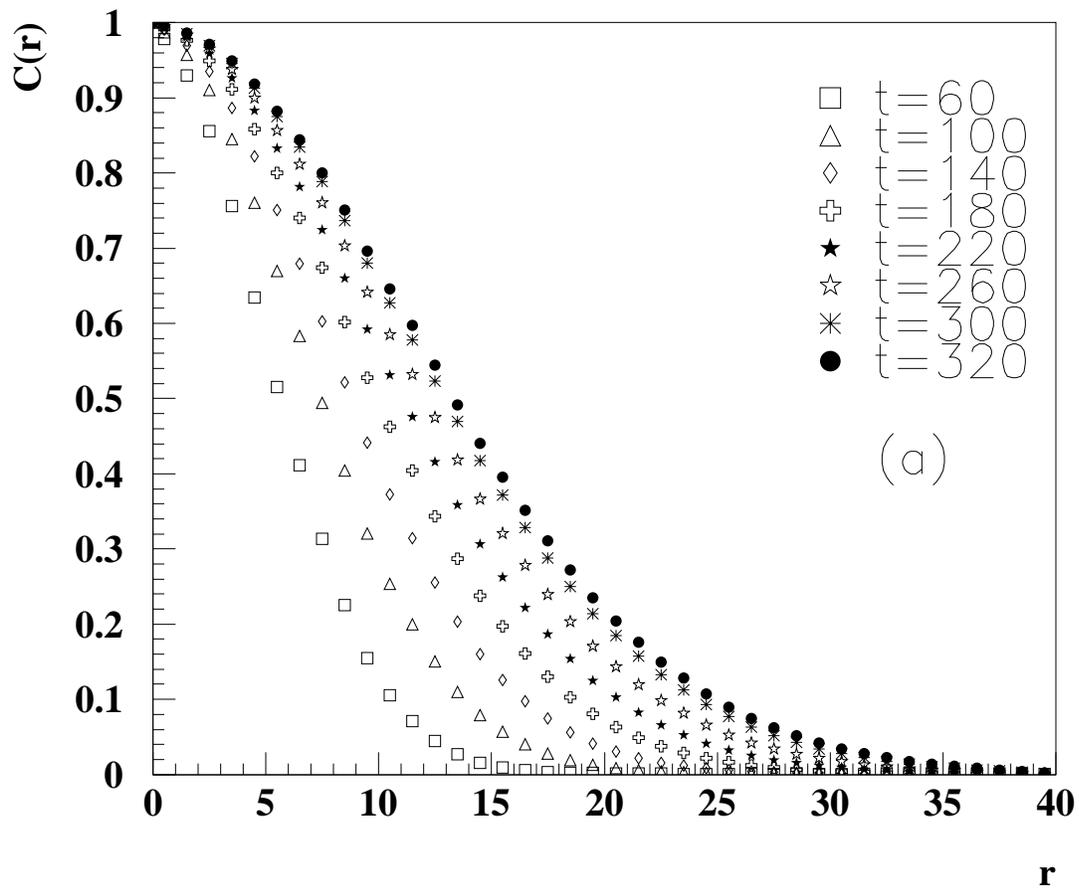
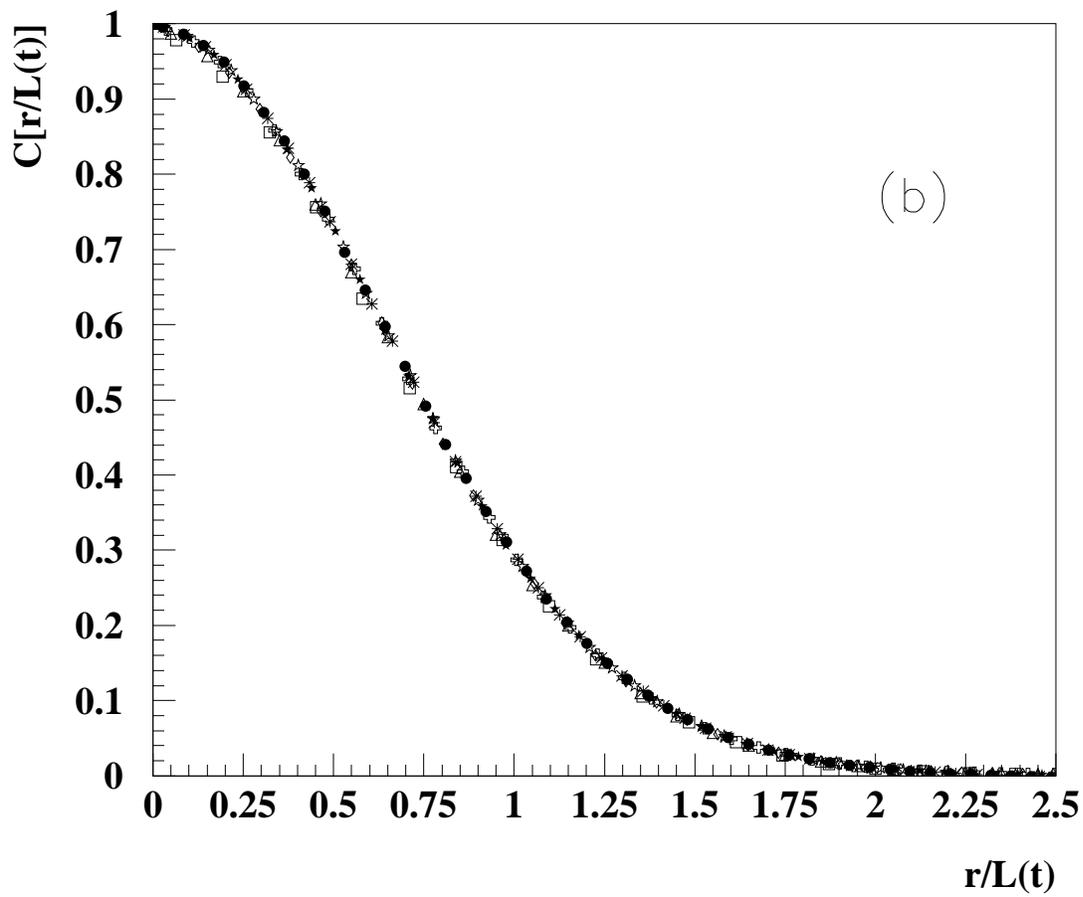
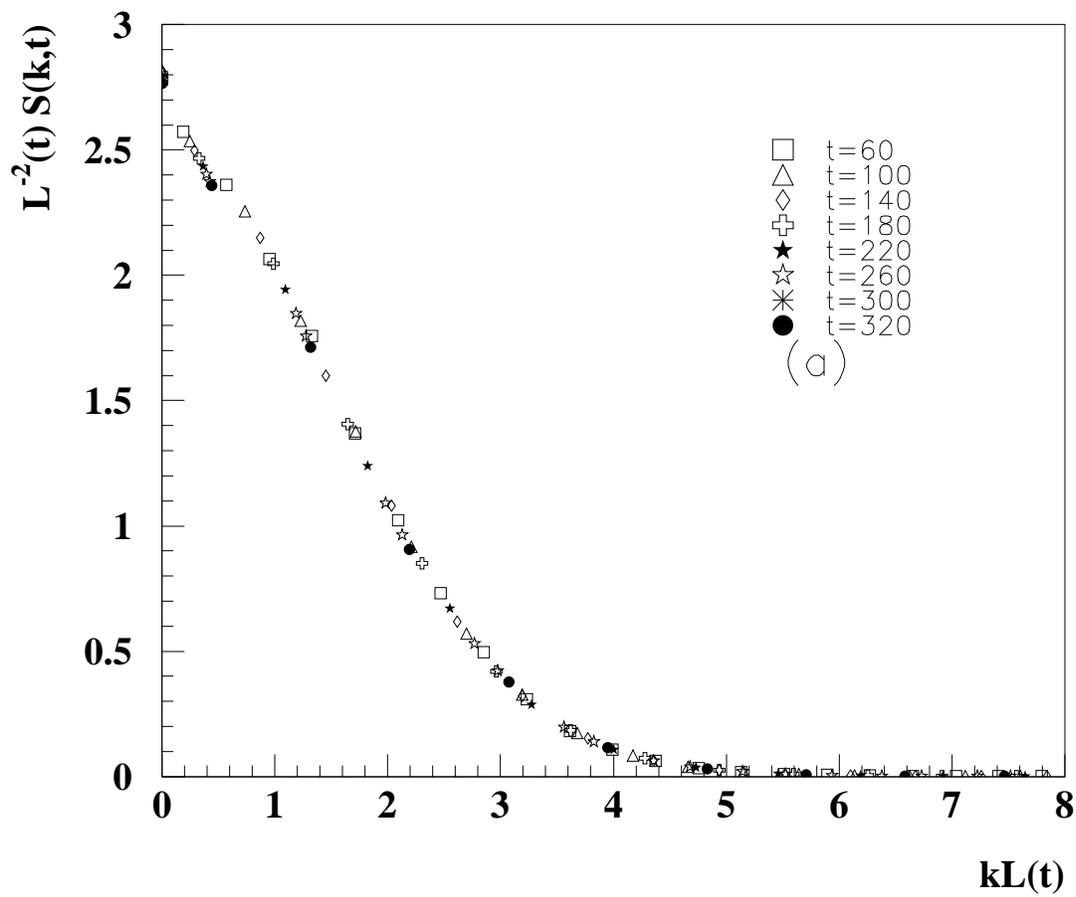
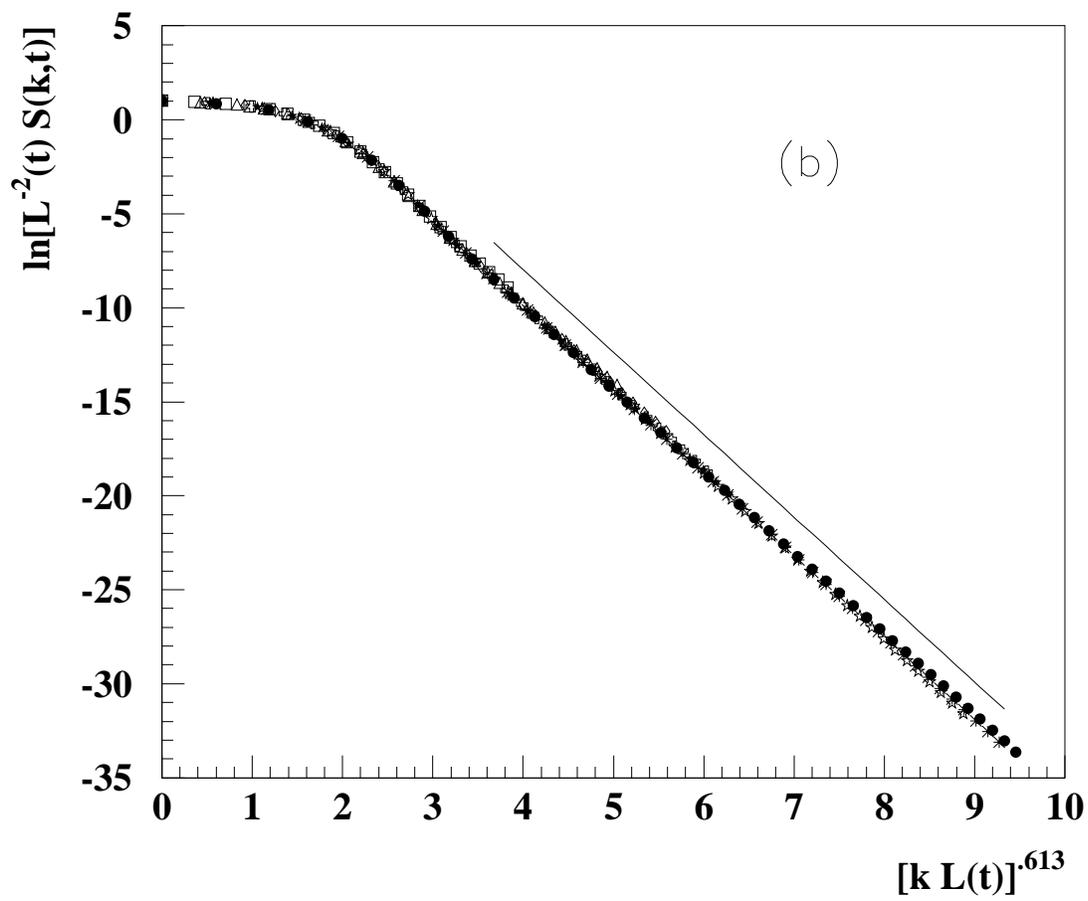
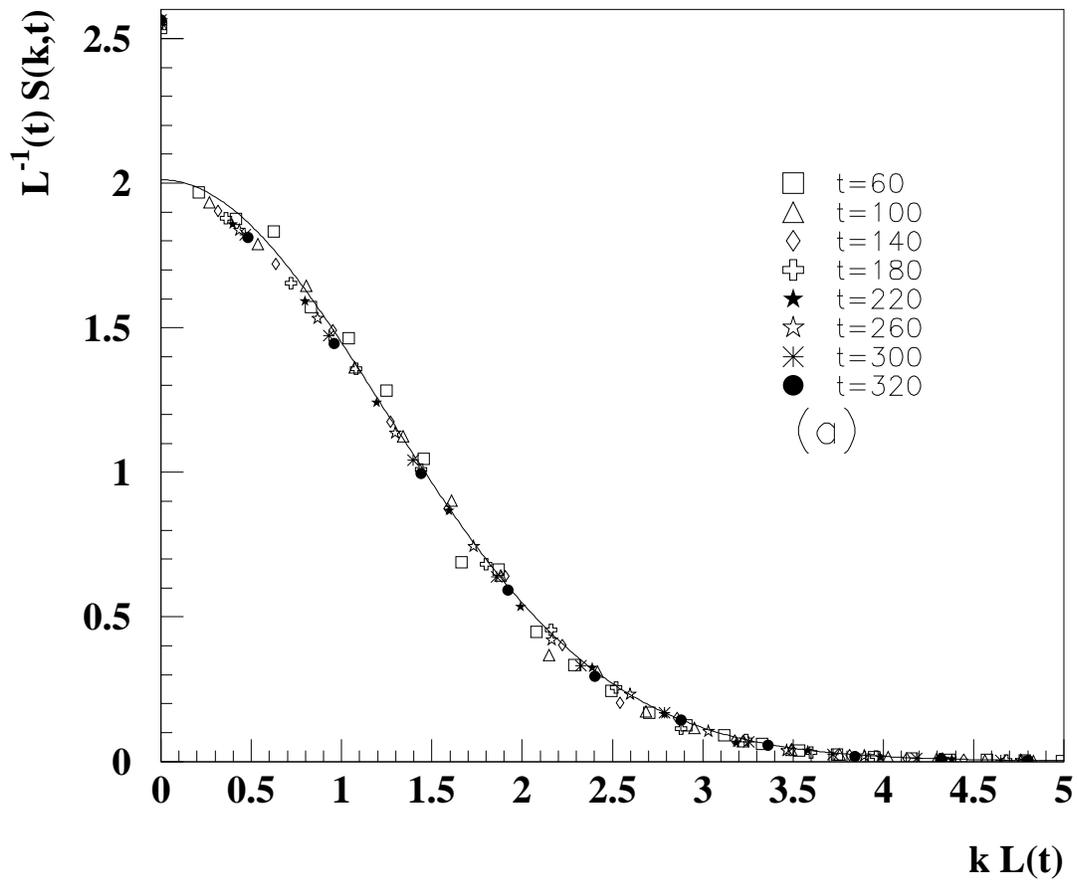
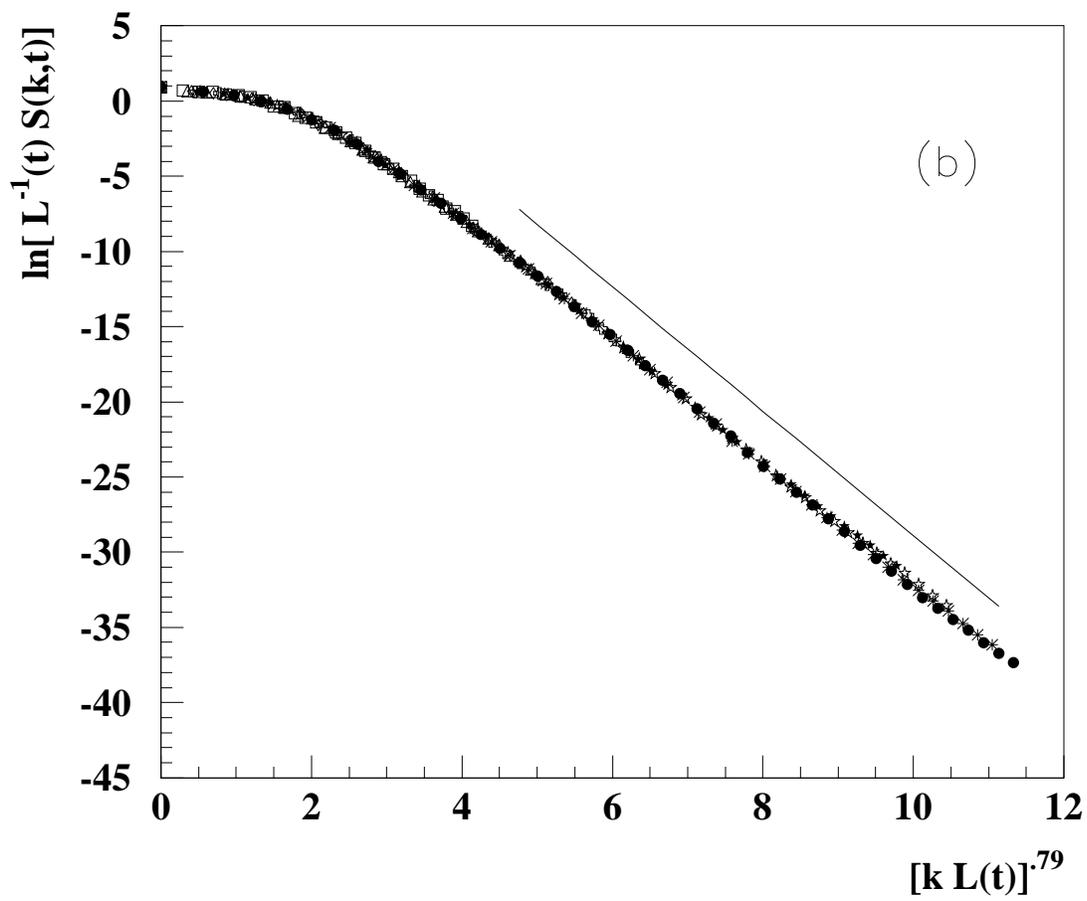
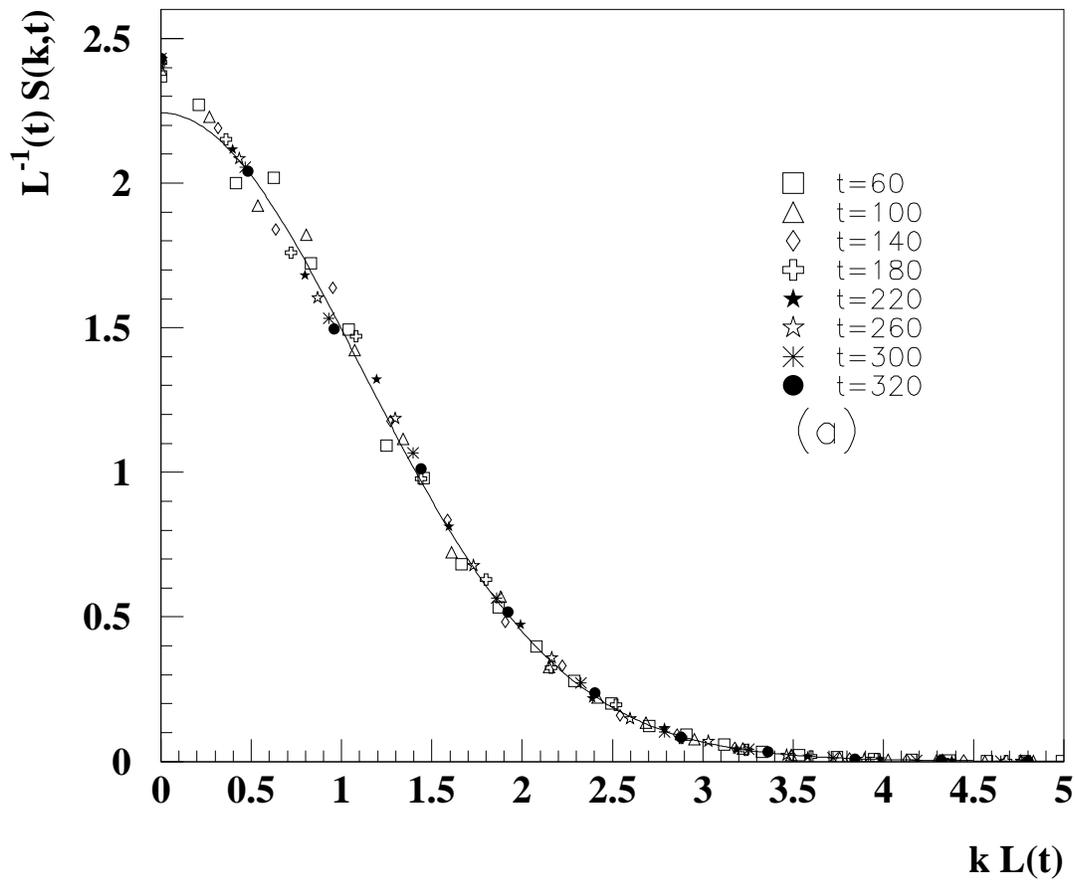
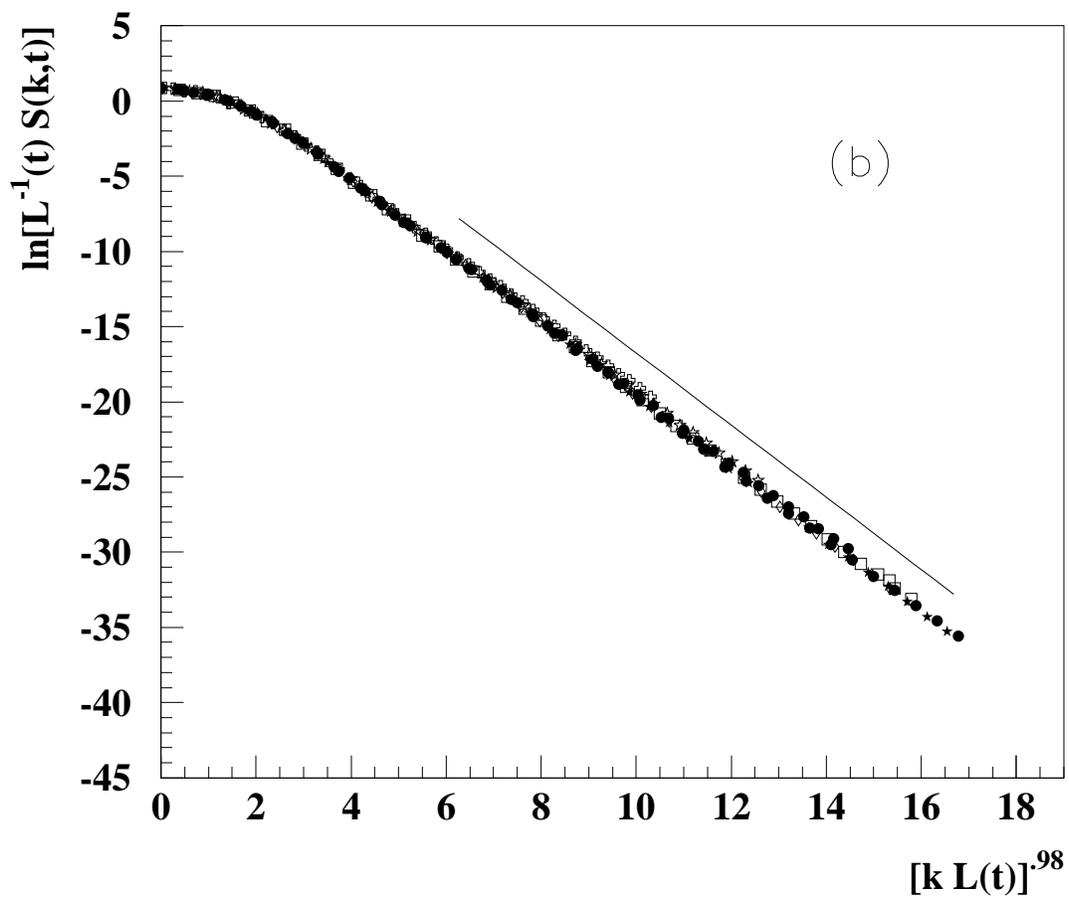
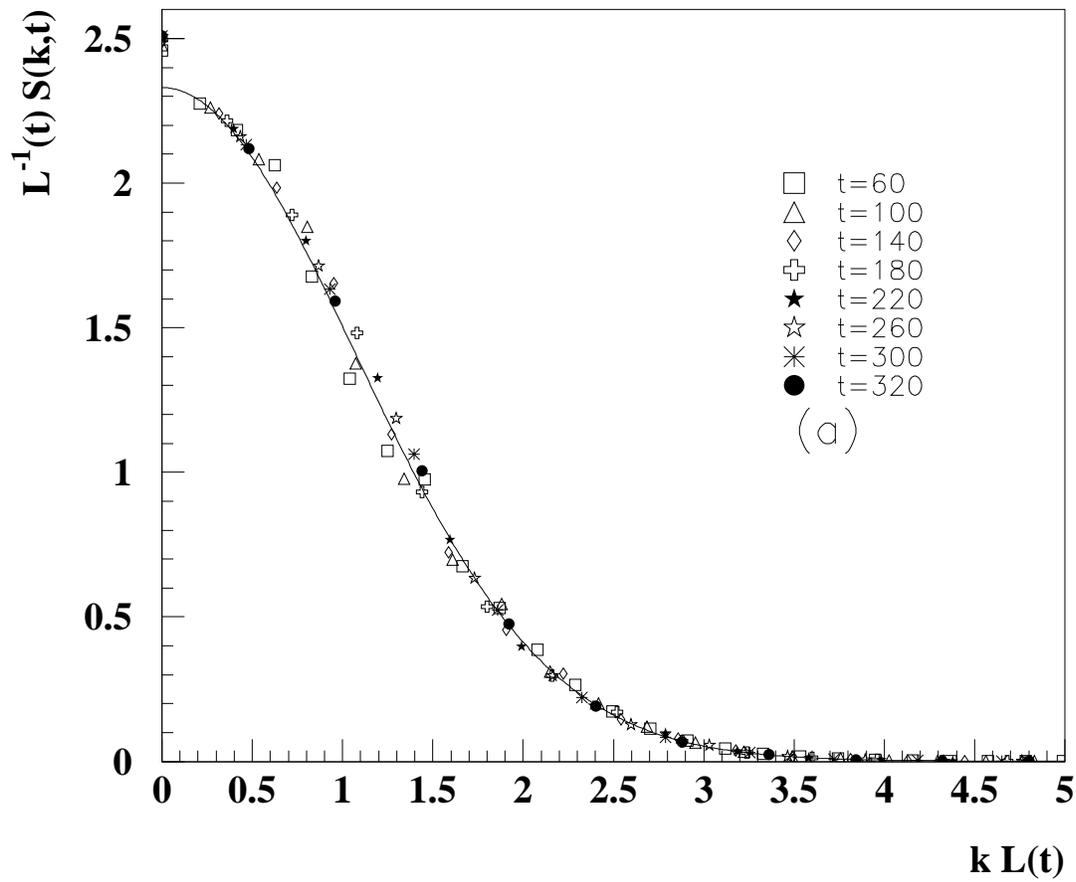
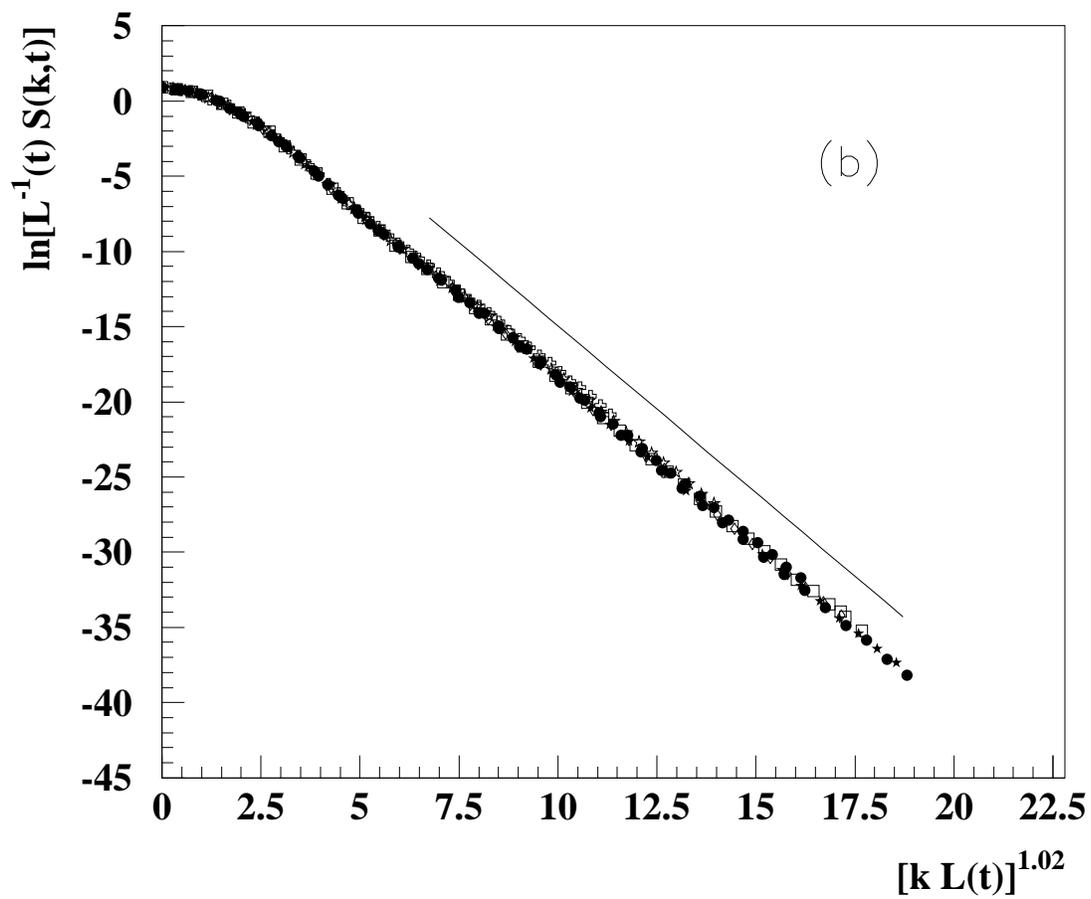
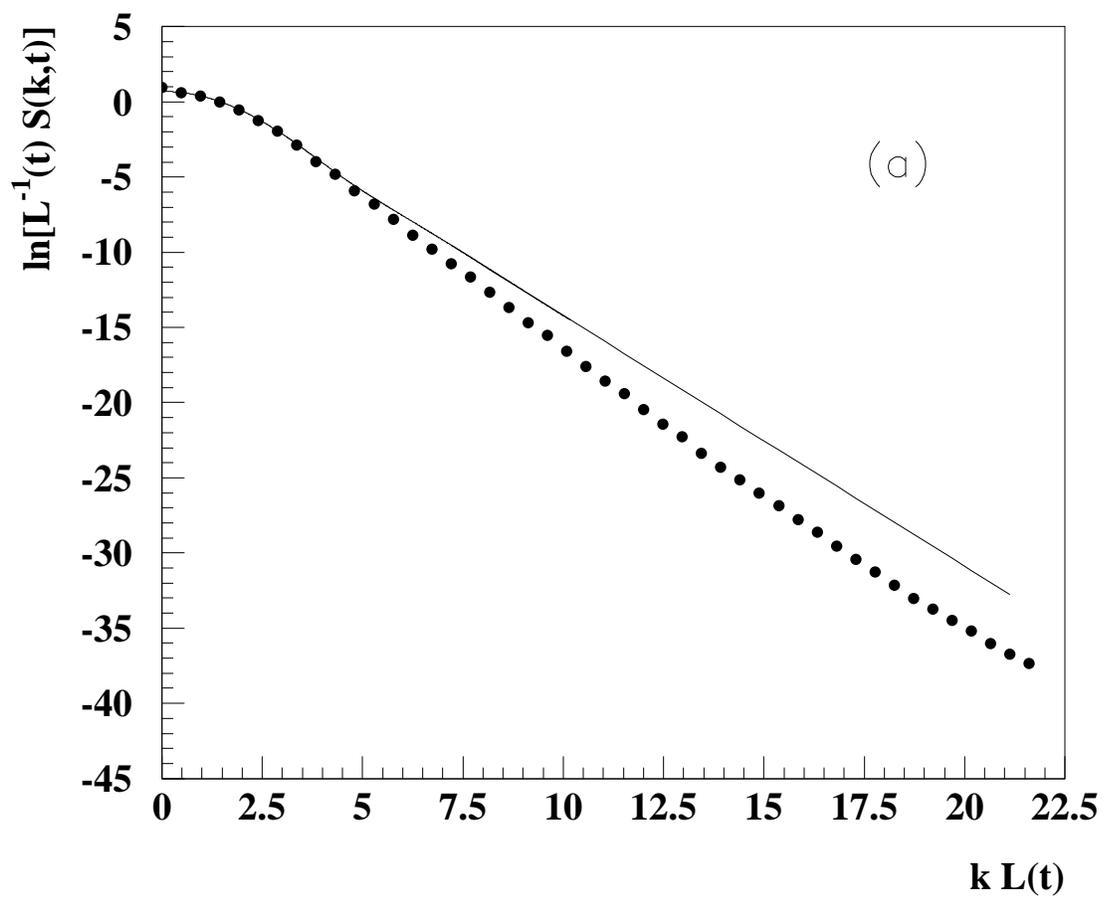
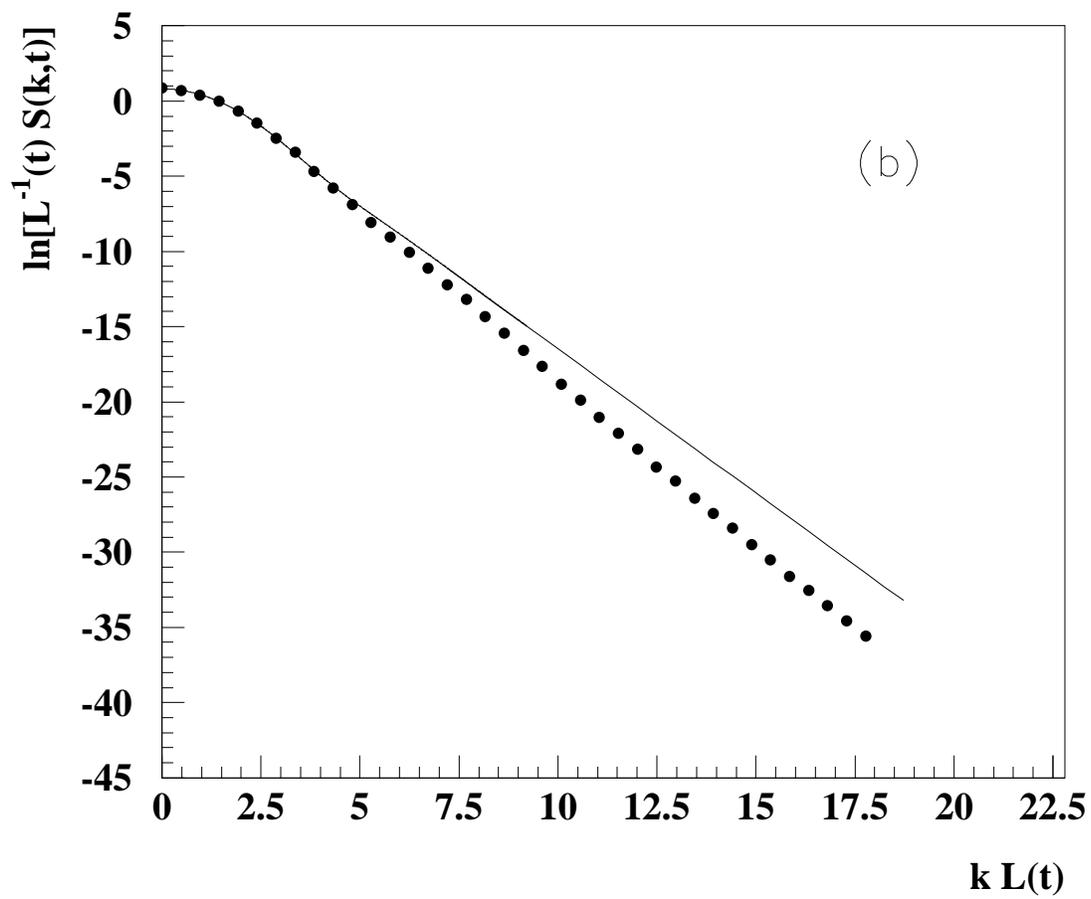
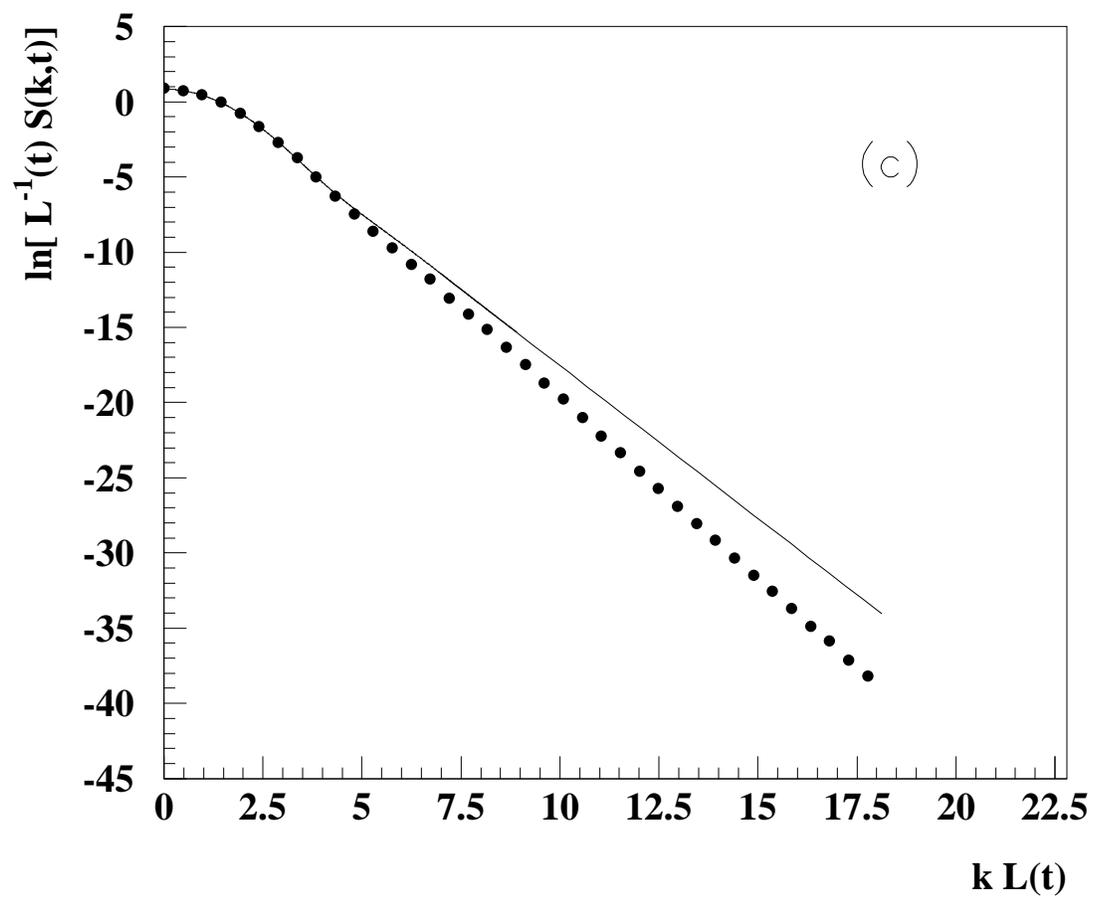
\end{document}